\documentclass[useAMS,usenatbib]{mn2e}
\usepackage{graphicx,amsmath}
\usepackage[usenames, dvipsnames]{color}

\title[Infall velocities near clusters of galaxies]{Infall near clusters of galaxies: comparing gas and dark matter velocity profiles}
\author[Albaek et al.]{L. Alb{\ae}k$^{1}$, 
S. H. Hansen$^{1}$\thanks{E-mail: hansen@dark-cosmology.dk}, 
D. Martizzi$^{2}$, 
B. Moore$^{3}$, 
R. Teyssier$^{3}$
\\
$^{1}$Dark Cosmology Centre, Niels Bohr Institute, University of Copenhagen,
  Juliane Maries Vej 30, 2100 Copenhagen, Denmark\\
$^{2}$Department of Astronomy and Theoretical Astrophysics Center, University of California, Berkeley, CA 94720-3411, USA.\\
$^{3}$Institute for Computational Science, University of Zurich, CH-8057 Zurich, Switzerland}
\begin{document}
\pagerange{\pageref{firstpage}--\pageref{lastpage}} \pubyear{2012}

\maketitle

\label{firstpage}

\begin{abstract}
We consider the dynamics in and near galaxy clusters.  Gas, dark
matter and galaxies are presently falling into the clusters between
approximately 1 and 5 virial radii.  At very large distances, beyond
10 virial radii, all matter is following the Hubble flow, and inside
the virial radius the matter particles have on average zero radial
velocity.  The cosmological parameters are imprinted on the infall
profile of the gas, however, no method exists, which allows a
measurement of it.  We consider the results of two cosmological
simulations (using the numerical codes RAMSES and Gadget) and find
that the gas and dark matter radial velocities are very similar.  We
derive the relevant dynamical equations, in particular the generalized
hydrostatic equilibrium equation, including both the expansion of the
Universe and the cosmological background.  {This generalized gas
  equation is the main new contribution of this paper.} We combine
these generalized equations with the results of the numerical
simulations to estimate the contribution to the measured cluster
masses from the radial velocity: inside the virial radius it is
negligible, and inside two virial radii the effect is below $40\%$,
{in agreement the earlier analyses for DM}.  We point out how the
infall velocity in principle may be observable, by measuring the gas
properties to distance of about two virial radii, however, this is
practically not possible today.
\end{abstract}


\section{Introduction}

The massive galaxy clusters are still in the process of accreting
material, so the gas, dark matter and galaxies just outside the virial radius are
presently infalling~\citep{1972ApJ...176....1G}.  This effect is most
visible when observing ongoing mergers, however, it may also be
visible in the smooth accretion of
material~\citep{2006AJ....132.1275R}. The details of the infall
profile depend on the cosmological
parameters~\citep{1974ApJ...193..525S,1989AJ.....98..755R,
  2013MNRAS.431.3319Z}, which makes it particularly interesting to
observe.

For dark matter this infall is already known to depend on the cluster mass
\citep{2006MNRAS.373.1409P, cuesta}, and this also implies that the
standard mass determination is affected outside the virial
radius~\citep{radek, falco}.  The gas is known to shock near the
virial radius, and at larger radii the gas is expected to be
free-falling onto the cluster together with the dark matter.
At distances beyond 5 virial radii, both gas and dark matter is
swept away with the Hubble expansion.

In this paper we will consider some of the details of the transition
region near the galaxy clusters. To this end we will consider the
results of 2 numerical cosmological simulations. We will compare the
dark matter and gas velocity profiles. We will also derive the
relevant equations (one for gas, and one for dark matter and galaxies)
including the effect of radial velocities. These equations turn out to
be {very} similar, despite the very different nature of the
particles involved and hence the different derivations.

Improving the mass profile reconstruction at and beyond the virial
radius is becoming relevant, as the sensitivity of Sunyaev-Zeldovich
and X-ray observations are becoming good enough to measure the gas
properties at these large radii. 
{Under the assumption of hydrostatic equilibrium, the observed
gas temperature and density gives the total mass profile
\citep{1986RvMP...58....1S}. However, magnetic fields, turbulence,
and other velocity terms like bulk motion and infall will induce extra
terms in the hydrostatic equilibrium equation, the socalled mass excess terms.}
Most other studies aiming at
improving the mass modelling of clusters focus on including
such non-thermal pressure components within the virial radius (see
e.g. \cite{2009ApJ...705.1129L,2009ApJ...691.1648F,2013ApJ...767...79S,2015MNRAS.448.1020S,2016MNRAS.455.2936S,2016arXiv160602293B})
or non-radially symmetric contributions
\citep{2012ApJ...758L..16S,2015MNRAS.448.1644S}.  \cite{rasia2006}
considered the bias on the hydrostatic equilibrium from the infall
velocities. 
Furthermore, \cite{rasia2006, rasia2012} used mock
catalogues to estimate the effect of the non-homogeneity of the
temperature on the hydrostatic equilibrium equation (HE).  The radial velocity
component described in this paper was most often not treated in a
consistent manner previously, mainly because the relevant equation
(which we derive here) has not been explicitly written down.

Below we will first derive the two generalized equations, one for gas
(generalizing the hydrostatic equilibrium) and one for the dark matter
(the generalized Jeans equation introduced in \cite{falco}). 
Analysing numerical simulations we
find that the infall velocity profiles for the gas and dark matter are
impressively similar. We then consider the effect on the mass
reconstruction, and demonstrate that the extra infall velocity term
contributes $20\%$ around 1.5 times the virial radius.
This paper will thus lay down the relevant equations, however, we will also
show that it is still not practically possible to measure the infall velocity
directly.

\section{Generalized hydrostatic equilibrium}

The Euler equation is the fluid equation representing conservation of momentum of a fluid
\citep{landau}
\begin{align*}
\frac{\partial \vec{v}}{\partial t} + (\vec{v} \cdot \nabla) \vec{v} = - \nabla \Phi - \frac{1}{\rho_{gas}} \nabla P \, .
\end{align*}
Here P and $\Phi$ are the gas pressure and total potential. 
Under the assumption of spherical symmetry, and using the ideal gas
law, the radial equation becomes
\begin{align}
\frac{\partial v_r}{\partial t} + v_r \frac{\partial v_r}{\partial r} = -\frac{k_b T_{gas}}{m_p \mu_{gas} r} (\frac{\partial ln \rho_{gas}}{\partial ln r} +\frac{\partial ln T_{gas}}{\partial ln r}) - \frac{\partial \Phi}{\partial r} \, ,
\label{b3}
\end{align}
where $v_r$ represents the radial velocity of a fluid particle.

The motion of any particle can at any radius be considered the sum of the Hubble
expansion and a peculiar velocity~\citep{1976ApJ...205..318P, 1978obco.meet....1G}
\begin{equation}
v_r = v_p + v_H \, ,
\end{equation}
where
\begin{align*}
v_H = H(t) r .
\end{align*}
For the gas particles, we will use the notation
\begin{equation}
v_r^{\rm gas} = v_{\rm gas} + v_H \, .
\end{equation}

The acceleration due to expansion is the time derivative of the Hubble law
\begin{align*}
\frac{d v_H}{d t} = - r H^2 q \, ,
\end{align*}
where $q$ is the deceleration parameter, $q=\Omega_m/2- \Omega_\Lambda$.

The background density must be included in the gravitational potential, whereby one 
gets~\citep{peebles,falco}
\begin{equation}
\frac{\partial \Phi (r)}{\partial r} = \frac{GM_{\rm tot}(r)}{r^2} + \frac{4\pi}{3}G\rho_b r +\frac{1}{3} \Lambda r \, ,
\end{equation}
where $\Lambda$ is the cosmological constant, $\Omega_\Lambda = \Lambda/3H^2$,
{and $M_{\rm tot}(r)$ is the total gravitating mass inside radius r.}
This can be rewritten as 
\begin{equation}
\frac{\partial \Phi}{\partial r} = \frac{G M_{\rm tot}(r)}{r^2} + r H^2 q \, .
\end{equation}

{After cancellation of a few terms the generalized Euler equation becomes}

\begin{eqnarray}
G M_{\rm tot}(r) &=& -\frac{k_b T_{gas} r}{m_p \mu_{gas}} (\frac{\partial ln \rho_{gas}}{\partial ln r} +\frac{\partial ln T_{gas}}{\partial ln r}) \nonumber \\
&&- r^2(\frac{\partial v_{\rm gas}}{\partial t} + H v_{\rm gas} + H r \frac{\partial v_{\rm gas}}{\partial r} + v_{\rm gas} \frac{\partial v_{\rm gas}}{\partial r}) \, , \nonumber \\ &&
\label{b5}
\end{eqnarray}
which can be written as
\begin{equation}
G M_{\rm tot}(r) = GM^{\rm HE}(r) - r^2 \tilde S(v_{\rm gas}) \, ,
\label{ghe}
\end{equation}
where $M^{\rm HE}(r)$ represents the standard hydrostatic equilibrium
terms. The extra term, $\tilde S(v_{\rm gas})$, vanishes for $v_{\rm gas}$ going to
zero.

\section{Generalized Jeans equation}
Dark matter and galaxies are treated as collisionless, and therefore
the fluid equations do not apply.  Instead one must start from the
collisionless Boltzmann equation.

By integrating over the velocities one obtains the Jeans equations.
\cite{falco,2013MNRAS.431L...6F} included both the expansion of the univese and the
background cosmology, and obtained the generalized Jeans equation
\begin{eqnarray}
G M_{\rm tot}(r) &=& - \sigma_r^2 r (\frac{\partial ln \rho}{\partial ln r} + \frac{\partial ln \sigma_r^2}{\partial ln r} + 2 \beta) \nonumber \\
&&- r^2 (\frac{\partial v_{\rm dm}}{\partial t} + H v_{\rm dm} + Hr \frac{\partial v_{\rm dm}}{\partial r} + v_{\rm dm} \frac{\partial v_{\rm dm}}{\partial r})\, ,  \nonumber \\ &&
\label{gje2}
\end{eqnarray}
{where $v_{\rm dm}$ is defined similarly to the gas peculiar velocity
$v_r^{\rm dm} = v_{\rm dm} + v_H$. The total gravitating mass, $M_{\rm tot}(r)$, includes both gas and DM, and 
the velocity anisotropy, $\beta$, measures the departure from an isotropic
velocity distribution of the DM velocities} 
\begin{equation}
\beta(r) \equiv 1 - \frac{\sigma_{\theta}^2 + \sigma_{\phi}^2}{2 \sigma_r^2} .
\end{equation}

The generalized Jeans equation can also be written as
\begin{equation}
GM_{\rm tot}(r) = GM^{\rm JE}(r) - r^2 S(v_{\rm dm}) .
\label{gje2b}
\end{equation}
where $M^{\rm JE}(r)$ represents the standard Jeans equation terms. The
extra term, $S(v_{\rm dm})$, vanishes for $v_{\rm dm}$ going to zero.

It is important to keep in mind, that $v_{\rm dm}$ here represents an
average velocity of the individual collisionless particles, which
differs from the fluid velocity $v_{\rm gas}$ in Eq.~(\ref{b5}). Despite the very different
derivations (and fundamentally different assumptions) the generalized
Jeans equation and the generalized hydrostatic equation look
remarkably similar.
{In the outer region where the gas collisions
are still not important the equations are naturally expected to look
similar since both equations essentially represent momentum
conservation. In the inner cluster region the similarity is more
surprising, since the collisionless DM has no equation of state, and
therefore the concept of pressure and temperature are not well defined
for the DM.}

\section{Numerical simulations}

In order to study the new terms in the generalized hydrostatic and
Jeans equations, we first need to find the peculiar velocity in and
near clusters. To this end we consider numerical simulations of
structure formation in a cosmological setting.

To have our results be fairly general, we chose to include two
different cosmological simulations, generated using both an Adaptive
Mesh Refinement (AMR) code, and a Smoothed Particle Hydrodamics (SPH)
code.  These represent two very different approaches to solving the
fluid dynamic equations of the gas component (see e.g. \cite{oscar,
  2012MNRAS.426.3112H}).

We are using two samples of simulated cluster haloes from
\cite{martizzi} and \cite{2011MNRAS.418.2234B}. We refer the reader
to those papers for details, but summarize the most important
properties of the simulations below for completeness.

The AMR code RAMSES \citep{teyssier} simulates structure formation
in a flat Universe with cosmological constant density parameter $\Omega_\Lambda
= 0.728$, matter density parameter $\Omega_m = 0.272$ of which the
baryonic density parameter is $\Omega_b = 0.045$, power spectrum
normalization $\sigma_8 = 0.809$, primordial power spectrum index $n_s
= 0.963$, and current epoch Hubble parameter $H_0 = 70.4
km/s/Mpc$. The simulation was initially run as a dark-matter-only
simulation with comoving box size 144 Mpc/h and particle mass $m_{dm}
= 1.55 \cdot 10^9 M_{\odot} /h$. Here h is the dimensionless Hubble
parameter, defined as $h = \frac{H_0}{100 km/s/Mpc}$. After the dark
matter only simulation was run, 51 cluster sized haloes with total
masses above $10^{14} M_{\odot}$ were identified. These regions were
then resimulated with a baryonic component, with dark matter particle
mass $m_{dm} = 1.62 \cdot 10^8 M_{\odot}/h$ and baryonic component
mass resolution of $3.22 \cdot 10^7 M_{\odot}$. The 51 resimulation
runs implemented models of radiation, gas cooling, star formation,
metal enrichment, super novae feedback and AGN feedback and were
evolved to the present day epoch. A detailed description of the
simulation can be found in \cite{martizzi}. From the resulting
catalogue of cluster haloes, profiles of various physical parameters
as functions of the normalized radius $\frac{r}{r_{200}}$ were
extracted for each of the 51 clusters, in radius from 0 to 2.0
$r_{200}$.  $r_{200}$ is defined as the radius inside which the
average density is 200 times $\rho_c$, the critical density of the
universe.

The SPH simulation was performed with the Gadget-3 SPH code
\citep{2005MNRAS.364.1105S}. The simulation was initially run as a
dark-matter-only simulation with $1024^3$ particles in a box of
comoving length 1 Gpc/h. It assumes a flat $\Lambda$CDM cosmology with
$\Omega_\Lambda = 0.76$, $\Omega_m = 0.24$,
$\sigma_8 = 0.8$, $n_s = 0.96$, and $H_0 = 72 km/s/Mpc$. From this
simulation 24 haloes were identified with masses over $10^{15}
M_{\odot}/h$, and 5 haloes centered on smaller systems.
These were then resimulated including gas physics,  $\Omega_b = 0.04$. In the
29 resimulations the mass of each dark matter particle was
$8.47 \cdot 10^8 M_{\odot}/h$ and the initial mass of each gas
particle was $1.53 \cdot 10^8 M_{\odot} /h$. The implemented physics
included radiation, gas cooling, chemical enrichment, star formation,
supernovae feedback as well as AGN feedback. The simulations were
evolved to the present day epoch. A detailed description of the
initial simulation can be found in \cite{2011MNRAS.418.2234B} while a
detailed description of the physics included in the resimulations can
be found in
\cite{2013MNRAS.436.1750R,2014MNRAS.438..195P,2013MNRAS.430.2638M}.
From the 29 resimulated clusters the same data was extracted as for
the RAMSES simulation.  The data of this simulation was averaged in
and extracted from shells linearly distributed according to the same
physical radii in different cluster haloes. For this reason the data
for different haloes does not extend to the same normalized radius.

Whereas most of the haloes from the Gadget-3 simulation have a total mass
above $10^{15} M_{\odot}$, only a few of the haloes from the RAMSES
simulation have a total mass above this threshold.

\begin{figure}
        \includegraphics[angle=0,width=0.49\textwidth]{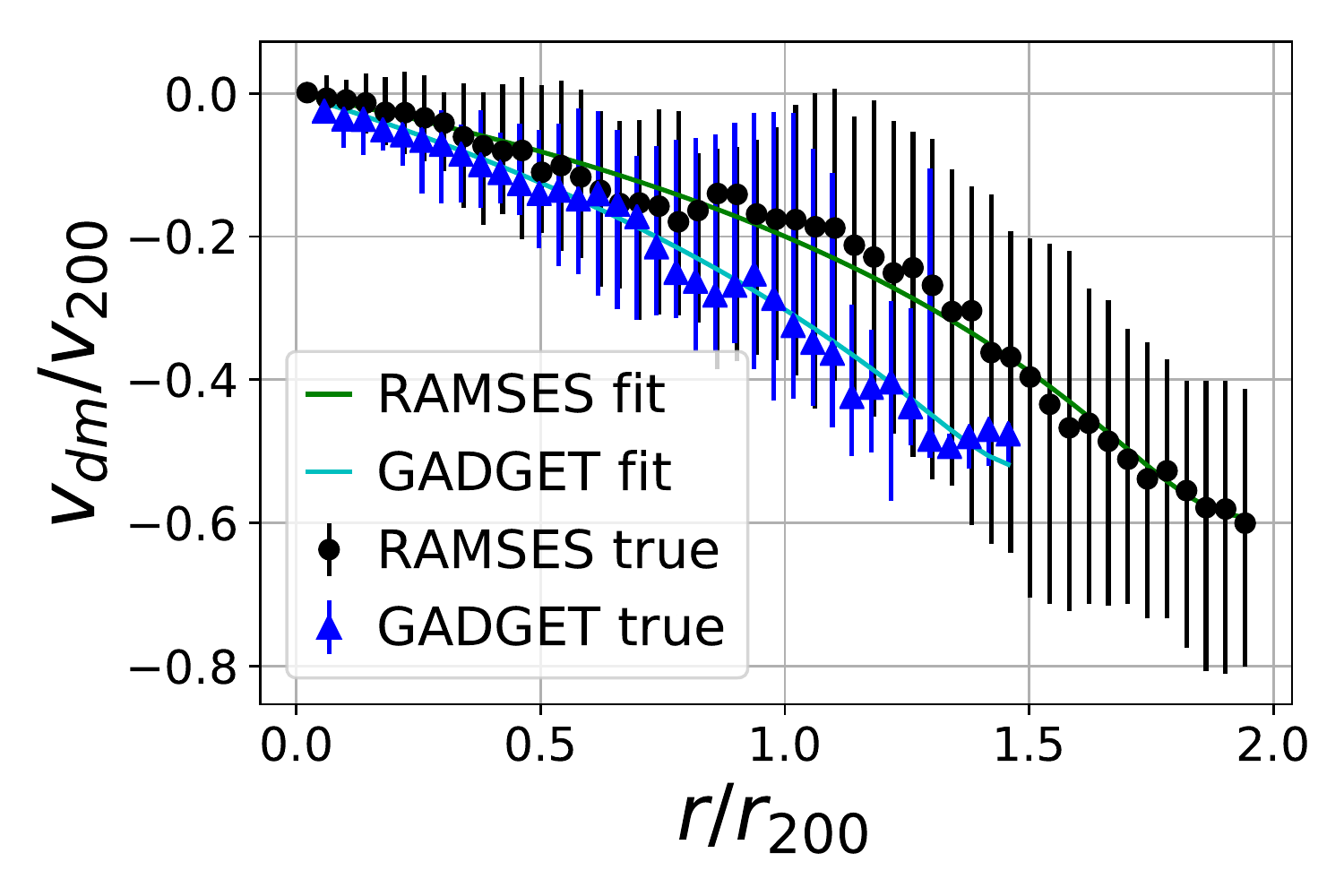}
\caption{The points represent the median of the normalized mean
  peculiar radial velocity, $\frac{v_{\rm dm}}{v_{200}}$, for the
  available clusters at the given radius. 
{The error bars represent $1\sigma$ sample variance between the many
simulated clusters. Each cluster has much smaller statistical error-bar,
however, the cluster-to-cluster variations lead to this large
dispersion.} The lines are numerical fits to the infall
  profiles, using the function described in section 5.}
\label{fig:fig1}
\end{figure}

\section{Peculiar velocity from numerical simulations}

The peculiar velocity of dark matter is shown in 
figure 1, as a function of radius.  The value of $v_{200}$ used to
normalize the profiles has been found for each cluster as
$\sqrt{\frac{G M_{200}}{r_{200}}}$.  
{The difference between the two simulations arises mainly from the
  different cluster masses considered, and to a much smaller extend
  from different cosmological parameters used.  For different
  cosmological models the turn around radius (where the average radial
  velocity is zero) changes, for instance for a larger cosmological
  constant the turn around radius is at smaller radii, as can be seen
  from the spherical collapse model \citep{2014JCAP...09..020P}.  For
  different cluster masses the detailed infall profile changes
  significantly \citep{2006MNRAS.373.1409P,cuesta}. Recently
  \cite{2016ApJ...832..185L} tried to quantify this effect, and
  fitted the peculiar velocities to the form suggested by
  \citep{falco}
\begin{equation}
v_{\rm dm} = -a \, \left( \frac{r}{r_{\rm vir}} \right)^{-b}  \, .
\end{equation}
They found that $b$ is about $0.26$ for structures of masses $\sim 4
\times 10^{13} M_\odot$, and increases to $b=0.43$ for masses above
$10^{14} M_\odot$. }

{In figure 2 we show the ratio between the DM and the gas infall
  profiles.  From this figure it is clear, that the both AMR and SPH
  simulations give very similar infall profiles for the gas and the
  DM. For a given cosmology and cluster mass, it is expected that the
  gas and DM infall profiles should agree outside the radius where gas
  cooling becomes important, as discussed in section 3.  The
  similarity of gas and DM infall profiles is expected to break down
  for smaller structures like galaxies with masses $\sim 10^{12}
  M_\odot$ as shown by \citep{2015ApJ...808...40W}. The gas cooling
  (and star formation and feedback) has very little effect for the
  large masses considered here, which is seen by the infall profiles
  essentially agreeing between the two different numerical simulations
  in the innermost regions.}

\begin{figure}
        \includegraphics[angle=0,width=0.49\textwidth]{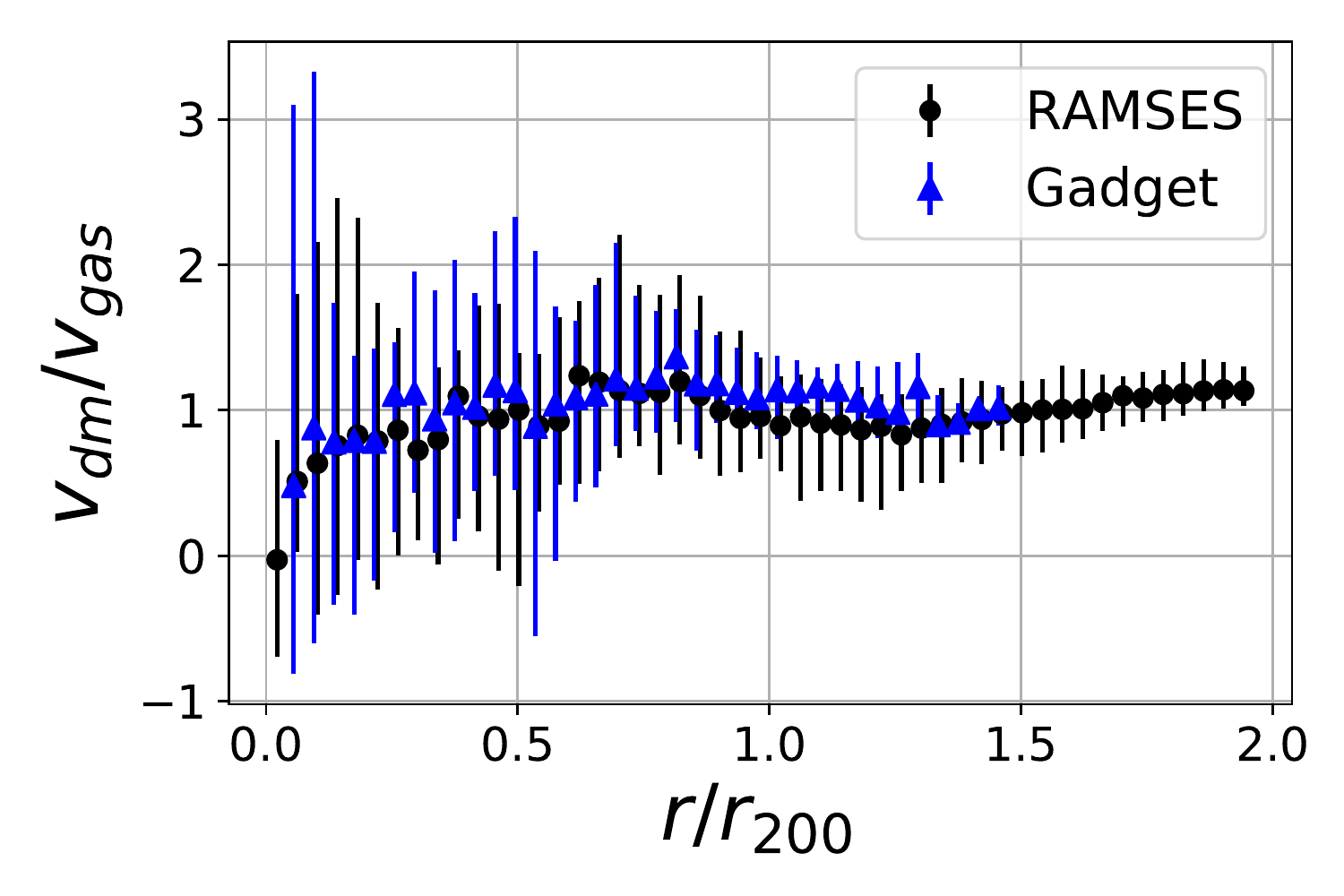}
\caption{The ratio of the DM to the gas infall profiles, showing that the gas
  and dark matter have virtually identical infall profiles.}
\label{fig:fig2}
\end{figure}

Having found the peculiar velocity, we can now consider the extra mass
terms in the generalized hydrostatic equilibrium and Jeans equations.

The two mass excess terms differ only in that $S(r,t)$ depends on the
average peculiar radial velocity of dark matter particles,
$v_{\rm dm}$, while $\widetilde{S}(r,t)$ depends on the peculiar
radial fluid velocity of the gas, $v_{\rm gas}$.

In order to calculate the mass excess terms, $S(r,t)$ and
$\widetilde{S}(r,t)$, $\frac{\partial v_{\rm dm}}{\partial t}$ and
$\frac{\partial v_{\rm dm}}{\partial r}$ must in principle be
determined for each simulated cluster.  In \cite{falco} the authors
used 7 different methods to attempt to estimate $\frac{\partial
  v_{\rm dm}}{\partial t}$. Each of the methods was either based
on theoretical calculations or on knowledge of the growth rate of
clusters in simulations. However, the results were not entirely
conclusive. 
Since the effect of $\frac{\partial v_{\rm dm}}{\partial t}$ is
expected to be very similar for the gas and the DM infall, we will
ignore that mass excess term for the present analysis. See appendix A
and B in \citep{falco} for an extended discussion on this point.

The mass excess therefore becomes simply

\begin{align}
S(r) \equiv  H_0 v_{\rm dm} + H_0 r \frac{\partial v_{\rm dm}}{\partial r} + v_{\rm dm} \frac{\partial v_{\rm dm}}{\partial r} ,
\label{massless}
\end{align}
with a similar definition for the ICM mass excess
$\widetilde{S}(r)$. Since the time evolution of S has been dropped,
and since both the simulations used are evolved to the
present epoch, the time dependent Hubble parameter $H(t)$ has been
reduced to its current value of $H_0$.

We chose to fit the infall profile.  In the central part one has
$\overline{v_r} = 0$, giving $v_{\rm dm} = -Hr$, so that the
fitting function should be linear as $r \to 0$.  At large radii there
is almost no braking due to the low density there, and all matter is
therefore in free fall towards the cluster center. The peculiar
kinetic energy of particles is thus simply the negative value of the
gravitational potential, $1/2 \, v_{\rm dm}^2 = - \Phi$, if they
are assumed to have fallen in from infinity. The density at this
radius is a tiny fraction of its central value, so the mass as a
function of radius is very nearly a constant $M(r) \approx
M_{total}$. This means that if the background density of the universe
is neglected, the potential is the Keplarian potential, $\Phi \approx
- \frac{G M_{tot}}{r}$. The peculiar radial velocity should therefore
be $v_{\rm dm} \approx - \sqrt{\frac{G M_{tot}}{r}} \propto
r^{-1/2}$, and the fitting function should then go as $r^{-1/2}$ in
the outer part. The precise shape of this part of the curve is not
crucial, as it is almost entirely outside the region that is
analyzed.

We use the form
\begin{eqnarray}
\frac{v_{\rm dm}}{v_{200}} &=& f\Big(\frac{r}{r_{200}}\Big) \nonumber \\
&=& - \alpha \frac{1}{\Big[(\frac{r}{r_{200}})^{-b} + C (\frac{r}{r_{200}})^{b/2} \Big]^{1/b} - D} \, ,
\label{martinafunc}
\end{eqnarray}
where $\alpha$, b, C, and D are positive parameters. This is a
slightly modified version of eq. (22) in \cite{falco}. It can be seen
that for $r \to 0$ the function approaches a linear shape, $f \to
-\alpha \frac{r}{r_{200}}$, while for $r \to \infty$ it goes to $f\to
-\frac{\alpha}{C} (\frac{r}{r_{200}})^{-1/2}$, as wished.  D
determines the shape of the function in the intermediate region where
braking occurs. The parameter $\alpha$ can either be chosen to act as
a free parameter, or be given the value $\alpha = \frac{H
  r_{200}}{v_{200}}$, for which the condition of a static central
region will be fulfilled.  {An example of such a fit {to the RAMSES simulated
profiles}
gives parameters $\alpha =0.137 \pm 0.0069, b=16.8 \pm 24, C=8.9
\times 10^{-8} \pm 1.6 \times 10^{-6}$ and $D=0.314 \pm 0.022$, 
  which in figure 1 is seen to provide an acceptable fit. 
{For the Gadget simulation the fit to $\alpha$ is approximately
$25\%$ larger, and the other parameters are witin the error-bars quoted
above.} The shown
fit has a $\chi^2$ of 1.11 with respect to the data, meaning that the
reduced $\chi^2$ is 0.25. The reason why the uncertainties on the
parameters b and c are so exceptionally large, is because they are
very strongly correlated in the inner part of the clusters, where data
is available. If further studies should choose to include the
free-falling outer regions of the clusters, then it should be possible
to break the degeneracy and better constrain these two parameters.}
{The turn-around is barely visible when considering only radii within
two virial radii, which leads the variable $C$ to be consistent with
zero. If we fix $C=0$, i.e. having no turn-around in velocity, then
the best fit parameters are $\alpha = 0.143$ and $D=0.289$ for the
RAMSES simulation.}

Using this fit, we can calculate the resulting mass excess terms in
eqs.~(\ref{ghe},\ref{gje2b}) which is shown in the top panel of Fig.~3.
Each of the 3 new terms in the generalized hydrostatic equilibrium and
generalized Jeans equations are shown in the lower panel of Fig.~3.

First, we see clearly that inside the virial radius, there is no
effect of the combination of Hubble flow and infall. At radii between
1 and 2 virial radii, we see that the radial velocity affects the mass
estimates up to $40\%$, both for gas and dark matter.

The mass excess of the dark matter found this way is in good agreement
with \cite{falco}, where the mass excess of the galaxy component in
simulated clusters was estimated and found to be between 20\% and 40\% 
inside 2 $r_{200}$.

\begin{figure}
        \includegraphics[angle=0,width=0.49\textwidth]{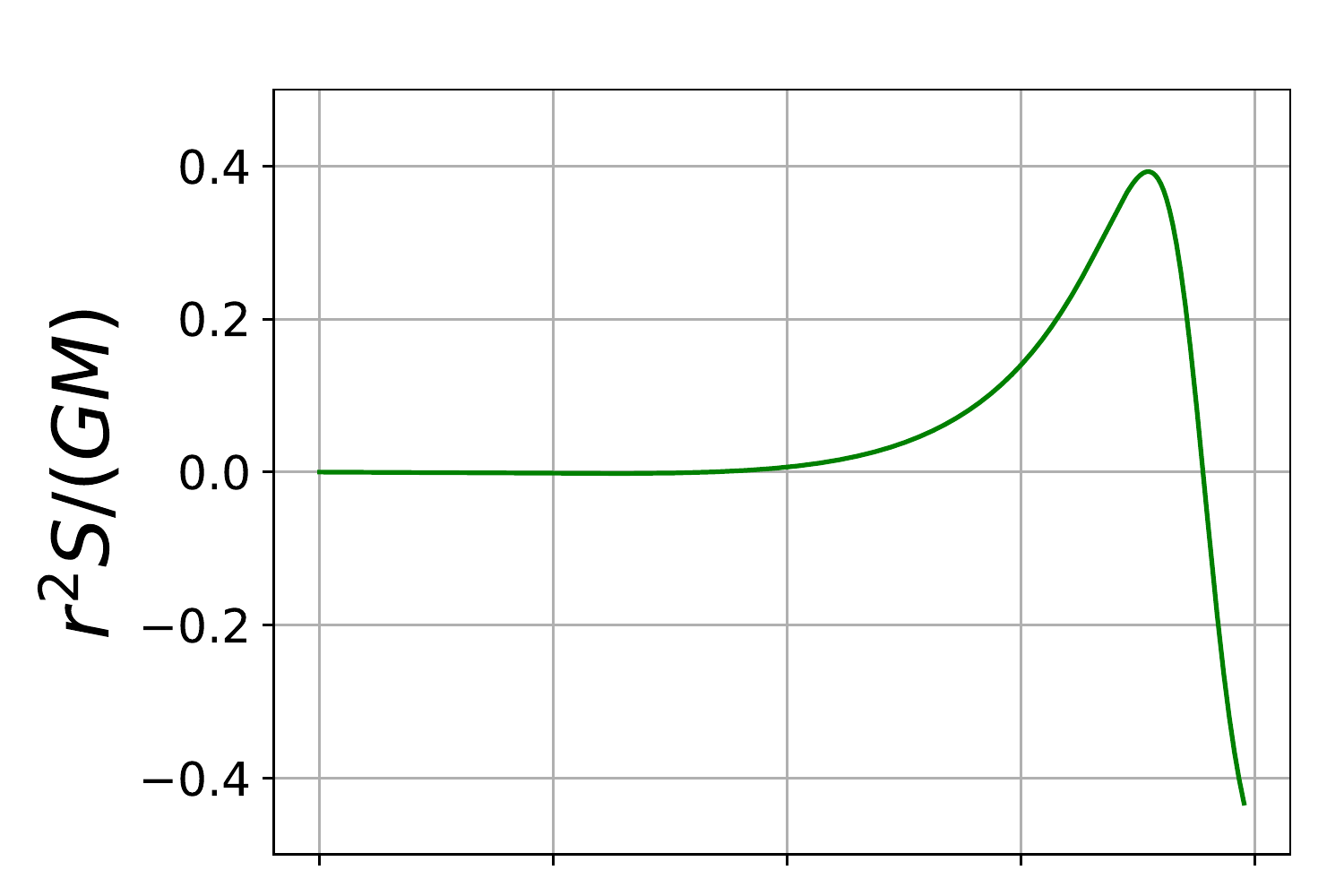}        
\includegraphics[angle=0,width=0.49\textwidth]{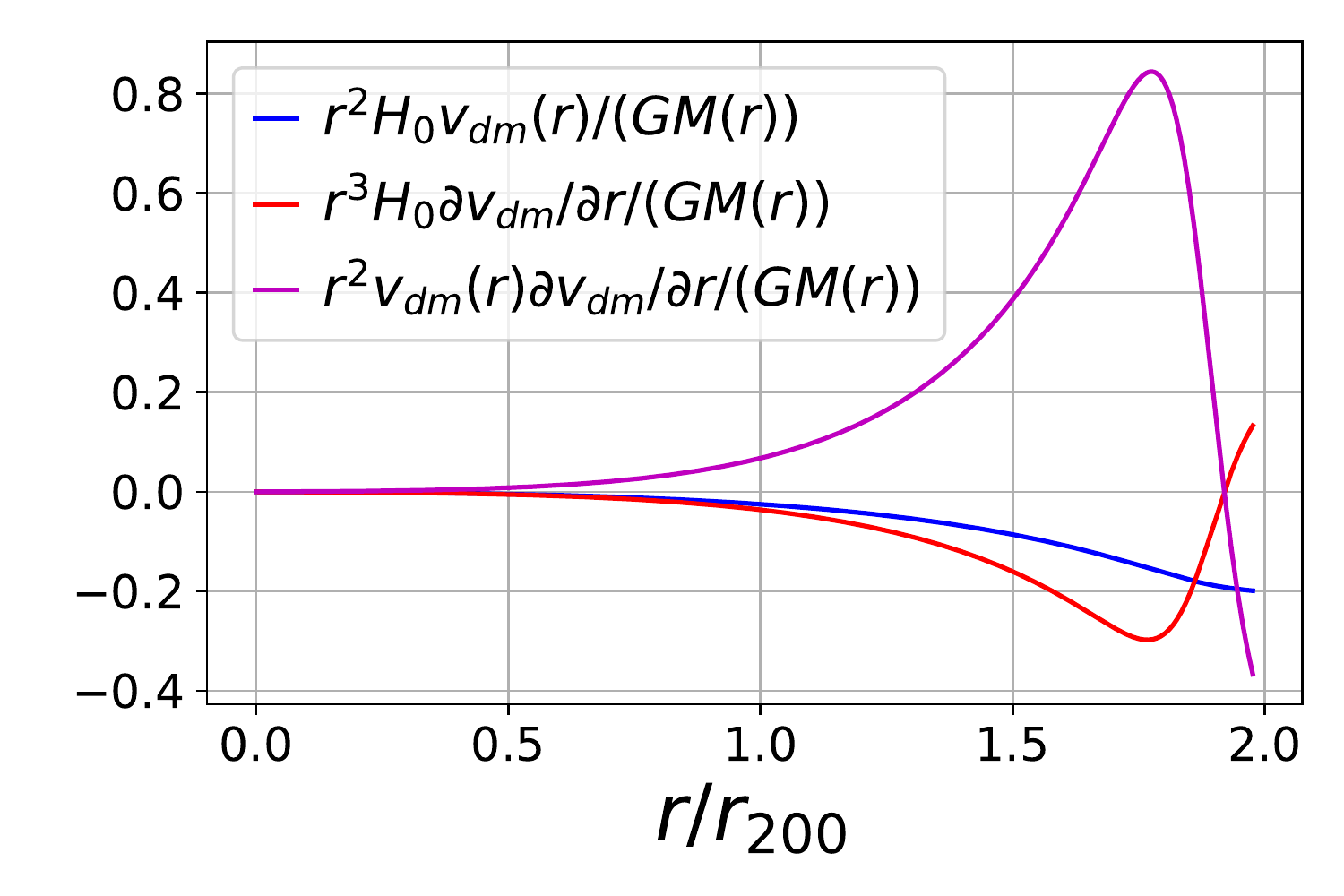}
\caption{The resulting mass excess divided by the total mass.  Upper
  panel: Inside the virial radius there is no effect of the
  combination of Hubble flow and infall. Upper panel: At radii between 1 and 2
  virial radii the mass estimates is systematically shifted up
  to $40\%$, both for gas and dark matter.  Lower panel: Each of the 3
  new terms in the generalized hydrostatic equilibrium.}
\label{fig:fig3}
\end{figure}

\section{Measuring the magnitude of the velocity infall}

{This technical section comes with the following warning: the section
will conclude that whereas it in principle would be possible to
measure a mass excess term coming from the infall profile, in practice
it is very difficult. Some readers may therefore prefer to go straight
to the concluding section.}

{It is fair to remind ourselves why we would be interested in
  measuring the magnitude of the infall velocity profile. The infall
  velocity is determined mainly by the cluster mass and cosmology, as
  described in section 5~(for earlier discussions on this point, see
  e.g. \cite{1974ApJ...193..525S,1989AJ.....98..755R,2013MNRAS.431.3319Z}). For
  instance the value of the cosmological constant will affect the
  position of the turn around radius, which is the radius where the
  infall velocity exactly cancells the Hubble expansion
  \citep{2014JCAP...09..020P}. Similarly, various modified gravity
  models have predictions for the magnitude of the infall velocity,
  which differs from those of $\Lambda$CDM
  \citep{2016arXiv161007268L}.  

The infall velocity leads to a mass
  excess term, as shown in eq.~(6).}  It is clear from the generalized
hydrostatic equilibrium, eq.~(\ref{b5}), that if one can
simultaneously measure accurately the total mass (for instance from
lensing) and gas density and temperature (for instance from either
X-ray or SZ), then one can directly get the mass excess term, and
hence estimate the {mass excess contribution from the velocity
  infall}.

This is, however, rather non-trivial, since a combination of
observational techniques always requires a careful control of
systematic effects. We will instead here entertain the possibility of
using only one method to measure gas parameters, and not include any
measure of the total mass.  The X-ray method is very well established
\citep{1986RvMP...58....1S}, however, it is also clear that the effect
of clumping may affect the precision of the mass determination, in
particular in the outer regions (see \cite{2015ApJ...806...43B} for a 
recent overview). Only few X-ray observations have today observed to the
virial radius \citep{urban}.

The SZ effect, on the other hand, is linear in both temperature and
density \citep{1972CoASP...4..173S}, which makes it possible to
measure the cluster temperature~\citep{1998A&A...336...44P,2002ApJ...573L..69H} without
concerns about clumpiness.  Future SZ observatories may in principle
deproject the spectra to get the full temperature and density
profiles~\citep{2004MNRAS.351L...5H}.

In principle the full velocity profile may be measured, but for
clarity we will here treat it as a one parameter search. This could be
the normalization $\alpha$ in eq.~(\ref{martinafunc}). We therefore
  consider the situation where the gas temperature and density have
  been measured accurately to large radii (e.g. two virial radii) and
  that $\alpha$ is unknown.

Let us clarify how to measure {the magnitude of the infall velocity}.
Pick a value of $\alpha$. From the generalized hydrostatic
equilibrium, eq.~(\ref{b5}) one can now calculate everything on the
r.h.s. This gives us the total mass, $M_{\rm tot}(r)$. By subtracting
the gas mass, we can now derive the dark matter density, which is one
of the parameters of the generalized Jeans equation,
eq.~(\ref{gje2}).

Using numerical simulations, the ratio of the gas temperature and the
dark matter ``temperature'' (i.e. the mean non-translational kinetic energy of dark matter particles) can be parametrized. 
This is a number which
today is believed to be fairly close to unity~\citep{2009ApJ...690..358H}, and future
numerical simulations will determine this ratio much more accurately.
Thereby we get $\sigma_r^2$ in the generalized Jeans equation,
eq.~(\ref{gje2}). The dark matter velocity anisotropy, $\beta$, is known to
  follow the density profile of the dark matter \citep{2006NewA...11..333H},
  or alternatively it can be parametrized from numerical
  simulations (see e.g. \cite{rasia2006}). 
  The last term is the mass excess term for the dark
  matter, but as we have found in this paper, it is virtually
  identical to the one of the gas, so with the assumed value for
  $\alpha$, it is known. One can therefore derive the total mass from
  the generalized Jeans equation, eq.~(\ref{gje2}).

We thus get the total mass in two different ways, and these can be
compared.  If they do not agree very well, then we consider a
different value for $\alpha$.  One loops over values of $\alpha$,
and then use some statistical optimization (like $\chi^2$).

It is worth repeating where the difference between the two derived
total mass profiles appears: if we consider a wrong value for
$\alpha$, then the derived mass from eq.~(\ref{b5}) is wrong, and
hence the derived density profile for the dark matter becomes
wrong. This is a rather subtle effect, and very high precision is
needed, both in level of equilibration, observation, and the
parametrized ratio of gas to dark matter ``temperatures''. Most likely
this will not be possible in the near future. The differences
between numerical techniques (AMR or SPH) still give a too large
systematic variation between the ratio of the dark matter and gas
temperature. 
In addition there are other known contributions to non-thermal
pressure (see the list of references on this issue in the introduction).
In a concrete implementation the measured infall
profile is therefore drowned in systematic error-bars. It therefore
makes little sense to implement the technique described above until
the origin of this difference between numerical simulation
techniques has been identified and clarified. The alternative might be
to include an independent measurement of the total mass, e.g. from lensing.

{We have hereby shown that whereas it in principle would be
  possible to measure a mass excess term coming from the infall
  profile, in practice it is very difficult. Furthermore, there will
  be other effects which also induce a mass excess: When using X-ray
  temperatures there is the problem of clumpiness, which could induce
  mass excess of the order $20\%$ near the virial radius
  \citep{2016ApJ...833..227A,2016arXiv161207260P}.  This problem could
  potentially be avoided by using the SZ effect to measure the
  temperature and density.  Furthermore, there could be bulk rotation
  or turbulence. These effects would be very difficult to remove from
  the data. And finally, the entire analysis gets even more convoluted
  (or impossible) when considering that structures usually depart from
  sphericity, see for instance the discussion in
  \cite{2012ApJ...748...21S,2016PASJ...68...97S,2016arXiv160302256V}.}

\section{Conclusion}

We compare the infall velocity near galaxy clusters from cosmological
simulations, and we find that the gas and dark matter infall profiles
are very similar. We derive the relevant gas equation, which is a
generalization of the hydrostatic equilibrium equation, and we find
that within two virial radii the infall velocity induces a mass excess
which is less than $40\%$. Inside the virial radius it is negligible.
{The similar generalized equation has previously been derived for
  collisionless DM}, and it effectively has an equivalent form.
We suggest how future detailed observations in principle may be used
to measure this infall profile. However, we also point out that the
precision of the needed ca\-li\-bration with numerical simulations is
still far too low for this method to be used in practice.

\section*{Acknowledgement}
It is a pleasure to thank Elena Rasia for providing the data
from the Gadget simulations. We thank Emiliano Munari and Elena Rasia
for comments on the manuscript.
We thank 
Christoffer Bruun-Schmidt, Beatriz Soret,
Catarina Fernandes and Joel Johansson for discussions.
Finally we thank the anonymous referee for good suggestions and
critical comments which improved the paper.
This project is partially funded by the Danish council for independent research under the 
project ``Fundamentals
of Dark Matter Structures'', DFF – 6108-00470.

\label{lastpage}

\end{document}